\documentstyle[aps,prb,floats,12pt]{revtex}
\def\beq{\begin{equation}}
\def\eeq{\end{equation}}

\def\P{{\tilde P}}
\def\psiT{{\psi_{\rm T}}}
\def\psiB{{\psi_{\rm B}}}
\def\ket#1{| {#1} \rangle}

\def\bra#1{ \langle{#1}|}

\def\psit#1{\psi_{{\rm T}#1}}
\def\psitbra#1{\bra{\psit #1}}

\def\psitket#1{\ket{\psit #1}}
\begin{document}
\tightenlines
%\draft
\title{ Universal Dynamics of Independent Critical Relaxation Modes}
\author{ M. P. Nightingale} 
\address{ Department of Physics, University of Rhode Island, Kingston RI
02881, USA}
\author{H.W.J. Bl\"{o}te }
\address { Department of Applied Physics, Delft University of Technology,\\
Lorentzweg 1, 2628 CJ Delft, The Netherlands }
%\date{\today}
\maketitle
\begin{abstract}
Scaling behavior is studied of several dominant eigenvalues of spectra
of Markov matrices and the associated correlation times governing
critical slowing down in models in the universality class of the
two-dimensional Ising model.  A scheme is developed to optimize
variational approximants of progressively rapid, independent
relaxation modes. These approximants are used to reduce the variance
of results obtained by means of an adaptation of a quantum Monte Carlo
method to compute eigenvalues subject to errors predominantly of
statistical nature.  The resulting spectra and correlation times are
found to be universal up to a single, non-universal time scale for
each model.

pacs {64.60.Ht, 02.70.Lq, 05.70.Jk, 64.60.Fr}
\end{abstract}

One of the most remarkable characteristics of critical behavior is
universality. For instance, it is generally accepted that upon
approach of a critical point, the correlation length diverges with a
power law and that the exponents are universal in the sense that they
depend only on the qualitative features characterizing the direction
of approach of the critical point. The definition of {\it the}
correlation length can be based on any of various correlation
functions, the most obvious ones of which are the order parameter and
energy autocorrelation functions.  The nature of the definition shows
up in the amplitude of the power law: Within a given universality
class, this amplitude is of the form $m A$, where $A$ is universal,
but depends on the observables used in the definition of the
correlation length; $m$ is non-universal, yet independent of these
detail, and a function only of the representative of the universality
class.

In terms of the spectrum of the transfer matrix one can define an
infinity hierarchy of correlation lengths. Indeed, the spectrum of the
transfer matrices, which has been studied extensively for
two-dimensional systems, displays such universal
amplitudes.\cite{TMreview} For statics the transfer matrix generates
translations in space.  The Markov matrix is its analog for
translations in time in for stochastic dynamics.  From this
perspective, the analog of the correlation length is the correlation
time and in this Letter we address universality of the critical point
amplitudes that describe the vanishing of gaps of the spectrum of the
Markov matrix and the slowing-down of the independent modes.

The dynamic critical exponent $z$ links the divergences of the
correlation length and time, $\xi$ and time $\tau$, via the relation
$\tau \sim \xi^z$. For a finite system of linear dimension $L$, with
its thermodynamic fields fixed at the critical values of the infinite
system, this becomes $\tau \sim L^z$, since the correlation length is
limited by the size of the system.  We denote by
$1=\lambda_{L0}>\lambda_{L1}\ge \dots$ the eigenvalues of the Markov
matrix, and by $\Delta_{Li}=1-\lambda_{Li}$ the spectral gaps. For the
associated correlation times one has $\tau_{Li}^{-1}= L^d\ln
\lambda_{Li}$ and the latter are expected to display the following scaling
\def\mt{{m_t}}
\beq
\tau_{Li} \approx L^{-d} \Delta_{Li}^{-1}  \to \mt A_i L^z,
\label{eq.mt}
\eeq
for $L\to \infty$. Here $A_i$ and $z$ are universal, and $\mt$ is a
non-universal, metric factor.\cite{NBua,Privman.Fisher.84,BNuniv}
Furthermore, $d$ is the dimensionality of the system, which enters
because we consider a Markov matrix that evolves the system in a local
sense only.  More specifically, we study single-spin-flip dynamics in
Ising models defined on a square lattice of linear dimension $L$ with
nearest and next-nearest neighbor couplings $K$ and $K'$ and periodic
boundaries.  We focus on models described by three ratios $\beta=
K'/K$, namely $\beta=-{1\over 4},0,1$, the {\it opposite-, nearest-},
and {\it equivalent-neighbor} models.

The dynamics is generated by the Markov matrix $P$, element $P(S',S)$
of which defines, given a configuration $S$, the conditional
probability of a transition to a configuration $S'$. If $S$ and $S'$
differ by more than one spin, $P(S',S)=0$. If both configurations
differ by precisely one spin,
\begin{equation}
P(S',S)= {1 \over L^2} \left\{ 
1 - \tanh \left[ \frac{ {\cal H}(S')-{\cal H}(S)}{2kT} \right] \right\},
\label{eq.markov}
\end{equation}
where ${\cal H}$ denotes the spin Hamiltonian.  The diagonal elements
$P(S,S)$ follow from the conservation of probability $\sum_{S'}
P(S',S)=1$.

We compute the spectrum of $P$ by means of a method used previously
for a single eigenstate\cite{NBdyn} generalized to several dominant
eigenvalues of the Markov matrix.  This method was introduced by
Ceperley and Bernu in the context of quantum Monte Carlo methods
\cite{CeperleyBernu88}.  A crucial element in our approach is the
construction of optimized trial states, and for this purpose we
generalize ideas of Umrigar {\it et al.}\cite{CyrusOptimization}.

The condition of detailed balance is used to define a stochastic
process that has the Boltzmann distribution $\exp[-{\cal H}(S)/kT]
\equiv \psi_{\rm B}(S)^2$ as its stationary state. Consequently,
$\P(S',S) \equiv \frac{1}{\psi_{\rm B}(S')} P(S',S)\psi_{\rm B}(S)$ is
symmetric in $S$ and $S'$.  We write eigenvectors of the transform
$\P$ in the functional form $\psi^{(\pm)}(S) \psiB(S)$ defined in
Eqs.~(12) and (13) of Ref.~\onlinecite{NBdyn}.  That is, in the first
place we restrict ourselves to translationally and rotationally modes,
which are even or odd under spin inversion, and, secondly, the trial
functions are written as linear combinations of monomials of the
magnetization and other long-wavelength Fourier transforms of the
spin configuration.

We generalize to simultaneous optimization of multiple trial states, a
powerful method \cite{CyrusOptimization,N96,NU96} of optimizing a
single many-parameter trial function. The latter is done by
minimization of the variance of the configurational eigenvalue:
Suppose that $\psiT(S,p)$ is the value of the trial function $\psiT$
for configuration $S$ and some choice of the parameters $p$ to be
optimized. The {\it configurational eigenvalue} $\lambda(S,p)$ of a
spin configuration $S$ is defined by
\beq
\psiT'(S,p) \equiv \lambda(S,p) \psiT(S,p), 
\label{eq.lamdas}
\eeq
where the prime indicates matrix multiplication by $\P$, {\it i.e.,}
$f'(S) \equiv \sum_{S'} \P(S,S') f(S')$ for arbitrary $f$.  The
optimal values of the variational parameters are obtained by
minimization of the variance of $\lambda(S,p)$, estimated by means of
a Monte Carlo sample.  We refer to Ref.~\onlinecite{NBdyn} for details
and mention only one of the key features of this method: in the ideal
case, {\it i.e.,} for an exact eigenstate $\psiT$, the variance
vanishes if it were to be computed exactly but also if one employs an
approximate Monte Carlo expression.  A similar zero-variance principle
holds for the method of simultaneous optimization of several trial
states to be discussed next.

For simplicity of presentation we first generalize the above method to
the more general ideal case in which one can exactly compute $m$
eigenvalues of the Markov matrix $\P$.  Suppose we have $m$ basis
states $\psit i$, $i=1,\dots,m$ and again $M$ spin configurations
$S_\alpha$, $\alpha=1,\dots,M$ sampled from $\psiB^2$.  The case we
consider is ideal in the sense that we assume that these states $\psit
i$ span an $m$-dimensional invariant subspace of $\P$.  In that case,
by definition there exists a matrix $\hat \Lambda$ of order $m$ such
that
\beq
\psit i'(S_\alpha)=\sum_{j=1}^m \hat\Lambda_{ij} \psit j(S_\alpha),
\label{eq.Lam_matrix}
\eeq
Again, the prime on the left-hand side of this equation indicates
matrix multiplication by $\P$.  If $M$ is large enough, $\hat\Lambda$
is for all practical purposes determined uniquely by the set of
equations (\ref{eq.Lam_matrix}) and one finds
\beq
\hat \Lambda=\hat N^{-1} {\hat {\cal P}}
\label{eq.Lam=P/N}
\eeq
where
\begin{eqnarray}
& \hat N_{ij}=Z^{-1}\sum_{\alpha=1}^M 
\psit i(S_\alpha) \psit j(S_\alpha) \nonumber \\
& {\hat {\cal P}}_{ij}=
Z^{-1} \sum_{\alpha=1}^M \psit i(S_\alpha) \psit j '(S_\alpha),
\label{eq.NP}
\end{eqnarray}
and where $Z$ is an arbitrary normalization constant; again, the prime
indicates matrix multiplication by $\P$.  In the non-ideal case, the
space spanned by the $m$ basis states $\psit i$ is not an invariant
subspace of the matrix $\P$.  In that case, even though
Eq.~(\ref{eq.Lam_matrix}) generically has no true solution,
Eqs.~(\ref{eq.Lam=P/N}) and (\ref{eq.NP}) still constitute a solution
in the least-squares sense, as may be verified by solving the normal
equations.

If a set of states span an invariant subspace, so does any
non-singular set of linear combinations.  In principle, the
optimization criterion should have the same invariance.  The spectrum
of the matrix $\hat \Lambda$ has this property, which suggests that
one subdivide the sample in subsamples and compute the variance of the
{\it local spectrum} over these subsamples.  In practice, however,
precisely this invariance gives rise to a near-singular non-linear
optimization problem.  Therefore, to avoid slow or no convergence, we
add a contribution to the above least-squares merit function to ensure
that the basis states {\em themselves} are good approximate
eigenstates, rather than just their linear combinations, and we use an
iterative optimization procedure: First a combination of the single
and multi-eigenstate merit functions is used, and finally the
resulting approximate eigenstates are optimized one at a time using
the single-state procedure only.  Unfortunately, this method is
capricious and often we proceed by trial and error.

The variational states can be used directly only to obtain results
with systematic errors, but these can be suppressed by the quantum
Monte Carlo projection method introduced by Ceperley and
Bernu\cite{CeperleyBernu88}. Define generalized matrix elements
\begin{eqnarray}
&N_{ij}(t)=\psitbra i \P^t \psitket j  \nonumber \\
&{\cal P}_{ij}(t)=\psitbra i \P^{t+1} \psitket j.
\label{eq.NPt}
\end{eqnarray}
For $t=0$, Eqs.~(\ref{eq.NP}) are Monte Carlo estimators for these
matrix elements, apart from the inconsequential normalization constant
$Z$.  One can view the matrix elements for $t>0$ as having been
obtained by the substitution $\ket{\psit i}\to {\P}^{t/2}\ket{\psit
i}$, which implies that spectral weights of ``undesirable'' states
lower down in the spectrum are reduced.  The matrix elements in
Eqs.~(\ref{eq.NPt}) are the following time auto- and cross-correlation
functions of the Markov process generated by the matrix $P$: $\langle
\psit i(S_0) \psit j(S_t)\rangle_{\psiB^2}$ and $\langle \psit i(S_0)
\psit j'(S_t)\rangle_{\psiB^2}$, where $S_0$ and $S_t$ are spin
configurations separated in time by $t$ single-spin flips.

It should be noted that in the limit of vanishing statistical error,
each eigenvalue estimate obtained by the above method is bounded from
above by the corresponding exact eigenvalue.  The reader is referred
to the work of Ceperley and Bernu in Ref.~\onlinecite{CeperleyBernu88}
for further details and references.  The systematic error decreases
for increasing projection time $t$ while the statistical error
increases.  An optimal intermediate $t$ has to be chosen, which yields
biased estimators and some uncertainty in the reliability of
statistical error estimates.

Of the three Ising-like models investigated here, the critical point
is exactly known only for the nearest-neighbor model, where it occurs
at $K=K_{\rm c}(0)= {1\over 2}\ln(1+\sqrt 2)$. The critical points of
the two crossing-bond models ---$K_{\rm c}(1)=0.1901926807(2)$ and
$K_{\rm c}(-{1\over 4})=0.6972207(2)$--- were determined by means of a
transfer-matrix technique combined with finite-size scaling
\cite{dynfs}. This analysis confirmed with a high precision that the
two crossing-bond models with belong to the static Ising universality
class.

Monte Carlo averages were taken over $1.2 \times 10^8$ spin
configurations, for system sizes in the range $5\leq L \leq 20$.  For
the nearest-neighbor model these samples were separated by a number of
Monte Carlo steps per spin equal to one for $L=5$ and increasing
quadratically to ten for $L=20$. For the other systems these numbers
where multiplied by the appropriate scale factors.  These surprisingly
short intervals are possible because the convergence of {\em the
eigenvalue estimates} as a function of projection time $t$ in
Eqs.~(\ref{eq.NPt}) is governed by lower-lying Markov matrix
eigenvalues.  These are much smaller than the largest odd eigenvalue,
which usually determines the relaxation rate.  For the system size
$L=5$, the Monte Carlo results for the largest odd eigenvalues of the
three models were compared with numerically exact results
\cite{NBdyn}. The consistency of both types of results confirms the
validity of our numerical procedures.

As noted before for the largest odd eigenvalue of the nearest-neighbor
model \cite{NBdyn}, the high statistical accuracy of the Monte Carlo
estimates of the eigenvalue is due to the accuracy of the
approximation of the eigenvector of the Markov matrix by the optimized
trial vectors.  The present Monte Carlo results for the largest odd
eigenvalues of the nearest-neighbor models agree with those of
Ref.~\onlinecite{NBdyn}. The new data are based on statistical sample
smaller by a factor of about 7, but the current trial vectors had more
variational freedom.

For finite system sizes $L$ we expect corrections to the leading scaling
behavior $\tau_L \sim L^z$. Following Ref.~\onlinecite{NBdyn}, we assume
corrections proportional to even powers of $1/L$:
\begin{equation}
\tau_{Li} \approx L^z \sum_{k=0}^{n_{\rm c}} \alpha_{ki} L^{-2k},
\label{fit}
\end{equation}
where the series is truncated at order $n_{\rm c}$. Although we cannot
exclude other powers in $1/L$, we have used Eq.~(\ref{fit}) to analyze
the Monte Carlo autocorrelation times.

Results of such fits with $n_{\rm c}=3$ are presented in Table
\ref{tab:z}.  The smallest systems do not fit Eq.~(\ref{fit}) well for
this value of $n_{\rm c}$. However, the residuals decrease rapidly
when $L_0$ the smallest system size included in the fit is
increased. The smallest acceptable value of $L_0$, as judged from the
$\chi^2$ criterion, is also included in Table \ref{tab:z}.

\begin{table}[htbp]
\caption{
Universality of the dynamic exponent $z$.  Results of least-squares
fits for the dynamic exponent for three Ising-like models and for five
distinct relaxation modes, identified in the first column: o$k$ refers
to odd mode number $k$ and e$k$ refers to the corresponding even mode.
Subsequent pairs of columns list $L_0$, the smallest system size
included in the fit, and the resulting estimates of $z(\beta)$ for
three ratios $\beta = K'/K$.  Estimated errors are shown in
parentheses.  To account for possible lack of convergence as a
function of projection time $t$ and flaws in Eq.~(\ref{fit}), two
standard errors are quoted.}
\vskip 1 ex
\begin{center}
\begin{tabular}{|c|c|l|c|l|c|l|}
  &$L_0$&$z(-1/4)$&$L_0$&$z(0)$&$L_0$&$z(1)$\\
\tableline
 o1   &   4   &2.163   (6)&   4   &2.1666 (14)&   4    &2.1659  (16) \\
 o2   &   5   &2.165   (6)&   6   &2.171   (4)&   8    &2.171   (4) \\
 o3   &   7   &2.11    (4)&   8   &2.178   (8)&   9    &2.167   (18) \\
 e2   &   6   &2.166   (6)&   5   &2.168   (2)&   5    &2.168   (2) \\
 e3   &   8   &2.17    (2)&   9   &2.14    (4)&   8    &2.19    (2)
\end{tabular}
\end{center}
\label{tab:z}
\end{table}

The estimates of $z$ obtained from the largest odd eigenvalues for the
three models shown in Table \ref{tab:z} are in a good agreement
mutually and also with the result $z=2.1665$ (12) of
Ref.~\onlinecite{NBdyn} for the nearest-neighbor model.  Universality
of $z$ has independently been confirmed by Wang and Hu \cite{Hu}, with
a level of precision in the order of $10^{-2}$.  The results for the
largest odd eigenvalues are in agreement with those obtained for the
other relaxation modes.  Although the differences do occasionally
amount to 3 $\sigma$, we attribute this to imperfections of
Eq. (\ref{fit}) and underestimation of the statistical errors of the
eigenvalues themselves.  Thus we interpret the data in Table
\ref{tab:z} as a confirmation of universality of the dynamic exponent
for different models and modes of relaxation.

\begin{figure}[htbp]
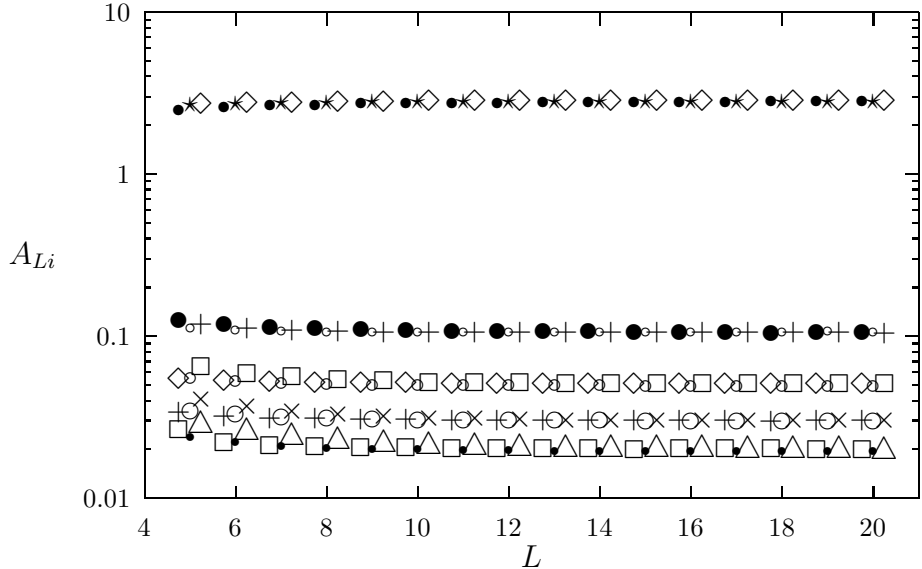

\begin{center}
\caption{
Universality of correlation-time amplitudes.  Semi-logarithmic plot of
the effective, size-dependent amplitudes $A_{Li}$. To separate data
point for the three models ${1\over 4}\mbox{sign} \beta$ was added to all
abscissae.  The data collapse predicted by Eq.~(\ref{eq.mt}) was
produced by fitting two metric factors, $\mt(0)$ and $\mt(-{1\over
4})$.  Amplitudes of odd and even states alternate in magnitude.}
\label{fig.scaled_gaps}
\vspace{4mm}
\input plot
\end{center}
\end{figure}

Correlation-time amplitudes were obtained from least-squares fits
using Eq.~(\ref{fit}) with $z$ fixed at $13\over6$, which happens to
be close to the most accurate results in Table \ref{tab:z}.
Subsequently, the non-universal metric factors $\mt$ were computed by
fitting to Eq.~(\ref{eq.mt}).  Defining $\mt(1)\equiv 1$, we found
$\mt(-{1\over 4})=2.391\pm 0.002$ and $\mt(0)= 1.5572 \pm 0.0005$.
Table \ref{tab:a} shows results of the fits.  Figure
\ref{fig.scaled_gaps} is a semi-logarithmic plot of the effective,
size-dependent amplitudes $A_{Li}(\beta)\equiv \tau_{Li} L^{-z}/\mt$
derived from the spectral gaps of the Markov matrices of the
opposite-, nearest-, and equivalent-neighbor Ising models,
$\beta=-{1\over 4}$, $0$ and $1$.

Finally, we note that if one suppresses all but the magnetization
dependence of the optimized trial functions, the number of nodes of
the resulting functions equals the number of the corresponding
eigenvalue counted from the top of the spectrum, which is in agreement
with the odd-even alternation shown in Tab.~\ref{tab:a}.

\begin{table}[htbp]
\caption{
Universality of correlation-time amplitudes. Results of least-squares
fits for the finite-size amplitudes for three Ising-like models and
for five distinct relaxation processes.  The first column and the ones
labeled $L_0$ are as in Table \ref{tab:z}. The columns labeled
$A_i(\beta)$ contain the amplitudes defined in Eq.~(\ref{eq.mt}) for
three interaction ratios $\beta = K'/K$ with metric factors $\mt$ as
given in the text.  Estimated errors, as defined in Table \ref{tab:z},
are shown in parentheses. The difference $A_i(1)-A_i(\beta)$ divided
by its error is denoted by $r$.}
\vskip 1 ex
\begin{center}
\begin{tabular}{|c|c|l|c|c|l|c|c|l|}
 &$L_0$&$A_i(-{1\over 4})$&$r$&$L_0$&$A_i(0)$&$r$&$L_0$&$A_i(1)$\\
\tableline
o1 & 5 & 2.827   (3) & 1.1& 5 & 2.8318  (8) &-0.6& 5 &2.8311 (10) \\
e2 & 6 & 0.10503 (2) & 0.1& 5 & 0.10504 (5) & 0.1& 5 &0.10504 (2) \\
o2 & 5 & 0.04970 (4) &-0.9& 6 & 0.04958 (2) & 1.6& 8 &0.04965 (4) \\
e3 & 6 & 0.03009 (5) & 0.3& 9 & 0.03013 (8) &-0.3& 8 &0.03011 (6) \\
o3 & 6 & 0.01956 (4) &-1.2& 8 & 0.01955 (4) &-0.9& 9 &0.01949 (4)
\end{tabular}
\end{center}
\label{tab:a}
\end{table}

This research was supported by the (US) National Science Foundation
through Grants DMR-9214669 and CHE-9625498 and by the Office of Naval
Research.  This research was conducted in part using the resources of
the Cornell Theory Center, which receives or received major funding
from the National Science Foundation (NSF) and New York State, with
additional support from the Advanced Research Projects Agency (ARPA),
the National Center for Research Resources at the National Institutes
of Health (NIH), IBM Corporation, and other members of the center's
Corporate Research Institute.

\end{document}